\documentclass[conference]{IEEEtran}
\IEEEoverridecommandlockouts
\makeatletter
\def\ps@headings{%
\def\@oddhead{\mbox{}\scriptsize\rightmark \hfil \thepage}%
\def\@evenhead{\scriptsize\thepage \hfil \leftmark\mbox{}}%
\def\@oddfoot{}%
\def\@evenfoot{}}
\makeatother
\usepackage{subfigure}
\usepackage{mathptmx}
\usepackage[scaled=-90]{helvet}
\usepackage{courier}
\usepackage[cmex10]{amsmath}
\usepackage{amsfonts,amssymb}
\usepackage{graphicx}
\usepackage{multirow}
\newtheorem{theorem}{Theorem}

\newtheorem{assumption}{Assumption}

\newcommand\T{\rule{0pt}{2.6ex}}
\newcommand\B{\rule[-1.2ex]{0pt}{0pt}}

\begin{document}

\title{SWIM: A Simple Model\\ to Generate Small Mobile Worlds\thanks{The work presented in this paper was partially funded by the FP7 EU project ``SENSEI, Integrating the Physical with the Digital World of the Network of the Future'', Grant Agreement Number 215923, www.ict-sensei.org.}}

\author{\IEEEauthorblockN{Alessandro Mei and Julinda Stefa}
\IEEEauthorblockA{Department of Computer Science\\
Sapienza University of Rome, Italy\\
Email: \{mei, stefa\}@di.uniroma1.it}}

\maketitle

\begin{abstract}
This paper presents small world in motion (SWIM), a new mobility model for ad-hoc networking. SWIM is relatively simple, is easily tuned by setting just a few parameters, and generates traces that look real---synthetic traces have the same statistical properties of real traces. SWIM shows experimentally and theoretically the presence of the power law and exponential decay dichotomy of inter-contact time, and, most importantly, our experiments show that it can predict very accurately the performance of forwarding protocols.
\end{abstract}


\section{Introduction}
\label{sec:intro}
Mobile ad-hoc networking has presented many challenges to the research community,
especially in designing suitable, efficient, and well performing protocols.
The practical analysis and validation of such protocols often depends on
synthetic data, generated by some mobility model. The model has the goal of
simulating real life scenarios~\cite{camp02wcmc} that can be used to tune
networking protocols and to evaluate their performance. A lot of work has been
done in designing realistic mobility models. Till a few years ago, the model of
choice in academic research was the random way point mobility model
(RWP)~\cite{rwp}, simple and very efficient to use in simulations.

Recently, with the aim of understanding human mobility~\cite{toronto, hui05,
hui06, milan07, UCAM-CL-TR-617}, many researchers have performed
real-life experiments by distributing wireless devices to people. From the data
gathered during the experiments, they have observed the typical distribution of
metrics such as inter-contact time (time interval between two successive
contacts of the same people) and contact duration. Inter-contact time, which
corresponds to how often people see each other, characterizes the opportunities
of packet forwarding between nodes. Contact duration, which limits the duration
of each meeting between people in mobile networks, limits the amount of data
that can be transferred.
In~\cite{hui05, hui06}, the authors show that the distribution of
inter-contact time is a power-law. Later, in~\cite{milan07}, it has been
observed that the distribution of inter-contact time is best described as a
power law in a first interval on the time scale (12 hours, in the experiments
under analysis), then truncated by an exponential cut-off. Conversely,
\cite{cai07mobicom} proves that RWP yields exponential inter-contact time
distribution. Therefore, it has been established clearly that models like RWP
are not good to simulate human mobility, raising the need of new, more realistic
mobility models for mobile ad-hoc networking.

In this paper we present small world in motion (SWIM), a simple mobility model that generates small worlds. The model is very simple to implement and very efficient in simulations. The mobility pattern of the nodes is based on a simple intuition on human mobility: People go more often to places not very far from their home and where they can meet a lot of other people. By implementing this simple rule, SWIM is able to raise social behavior among nodes, which we believe to be the base of human mobility in real life.
We validate our model using real traces and compare the distribution of inter-contact time, contact duration and number of contact distributions between nodes, showing that synthetic data that we generate match very well real data traces.
Furthermore, we show that SWIM can predict well the performance of forwarding protocols. We compare the performance of two forwarding protocols---epidemic forwarding~\cite{vahdat00epidemic} and (a simplified version of) delegation forwarding~\cite{dfw08}---on both real traces and synthetic traces generated with SWIM. The performance of the two protocols on the synthetic traces accurately approximates their performance on real traces, supporting the claim that SWIM is an excellent model for human mobility.

The rest of the paper is organized as follows: Section~\ref{sec:relatedwork} briefly reports on current work in the field; in Section~\ref{sec:solution} we present the details of SWIM and we prove theoretically that the distribution of inter-contact time in SWIM has an exponential tail, as recently observed in real life experiments; Section~\ref{sec:experiments} compares synthetic data traces to real traces and shows that the distribution of inter-contact time has a head that decays as a power law, again like in real experiments;
in Section~\ref{sec:forwarding} we show our experimental results on the behavior of two forwarding protocols on both synthetic and real traces; lastly, Section~\ref{sec:conclusions} present some concluding remarks.

\section{Related work}
\label{sec:relatedwork}

The mobility model recently presented in~\cite{levy} generates movement traces
using a model which is similar to a random walk, except that the flight lengths
and the pause times in destinations are generated based on Levy Walks, so with
power law distribution. In the past, Levy Walks have been shown to approximate
well the movements of animals. The model produces inter-contact time
distributions similar to real world traces. However, since every node moves
independently, the model does not capture any social behavior between nodes.
In~\cite{musolesi07}, the authors present a mobility model based on social
network theory which takes in input a social network and discuss the community
patterns and groups distribution in geographical terms. They validate their
synthetic data with real traces and show a good matching between them.

The work in \cite{LCA-CONF-2008-049} presents a new mobility model for clustered networks. Moreover, a closed-form expression for the stationary distribution of node position is given. The model captures the phenomenon of emerging clusters, observed in real partitioned networks, and correlation between the spatial speed distribution and the cluster formation.

In~\cite{workingDay}, the authors present a mobility model that simulates the every day life of people that go to their work-places in the morning, spend their day at work and go back to their homes at evenings. Each one of this scenarios is a simulation per se. The synthetic data they generate match well the distribution of inter-contact time and contact durations of real traces.

In a very recent work, Barabasi et al.~\cite{barabasi08} study the trajectory of
a very large (100,000) number of anonymized mobile phone users whose position is
tracked for a six-months period. They observe that human trajectories show a
high degree of temporal and spatial regularity, each individual being
characterized by a time independent characteristic travel distance and a
significant probability to return to a few highly frequented locations. They
also show that the probability density function of individual travel distances
are heavy tailed and also are different for different groups of users and
similar inside each group. Furthermore, they plot also the frequency of visiting
different locations and show that it is well approximated by a power law. All
these observations are in contrast with the random trajectories predicted by
Levy flight and random walk models, and support the intuition behind SWIM.

\section{Small World in Motion}
\label{sec:solution}

We believe that a good mobility model should
\begin{enumerate}
	\item be simple; and
	\item predict well the performance of networking protocols on real mobile
	networks.
\end{enumerate}
We can't overestimate the importance of having a \emph{simple} model. A simple
model is easier to understand, can be useful to distill the fundamental
ingredients of ``human'' mobility, can be easier to implement, easier to tune
(just one or few parameters), and can be useful to support theoretical work. We
are also looking for a model that generates traces with the same statistical
properties that real traces have. Statistical distribution of inter-contact time
and number of contacts, among others, are useful to characterize the behavior of
a mobile network. A model that generates traces with statistical properties that
are far from those of real traces is probably useless. Lastly, and most
importantly, a model should be accurate in predicting the performance of network
protocols on real networks. If a protocol performs well (or bad) in the model,
it should also perform well (or bad) in a real network. As accurately as
possible.

None of the mobility models in the literature meets all of these properties. The random way-point mobility model is simple, but its traces do not look real at all (and has a few other problems). Some of the other protocols we reviewed in the related work section can indeed produce traces that look real, at least with respect to some of the possible metrics, but are far from being simple. And, as far as we know, no model has been shown to predict real world performance of protocols accurately.

Here, we propose \emph{small world in motion} (SWIM), a very simple mobility
model that meets all of the above requirements. Our model is based on a couple
of simple rules that are enough to make the typical  properties of real traces
emerge, just naturally. We will also show that this model can predict the
performance of networking protocols on real mobile networks extremely well.

\subsection{The intuition}

When deciding where to move, humans usually trade-off. The best supermarket or the most popular restaurant that are also not far from where they live, for example. It is unlikely (though not impossible) that we go to a place that is far from home, or that is not so popular, or interesting. Not only that, usually there are just a few places where a person spends a long period of time (for example home and work office or school), whereas there are lots of places where she stays less, like for example post office, bank, cafeteria, etc. These are the basic intuitions SWIM is built upon. Of course, trade-offs humans face in their everyday life are usually much more complicated, and there are plenty of unknown factors that influence mobility. However, we will see that simple rules---trading-off proximity and popularity, and distribution of waiting time---are enough to get a mobility model with a number of desirable properties and an excellent capability of predicting the performance of forwarding protocols.

\subsection{The model in details}

More in detail, to each node is assigned a so called \emph{home}, which is a
randomly and uniformly chosen point over the network area. Then, the node itself
assigns to each possible destination a \emph{weight} that grows with the
popularity of the place and decreases with the distance from home. The weight
represents the probability for the node to chose that place as its next
destination.

At the beginning, no node has been anywhere. Therefore, nodes do not know how
popular destinations are. The number of other nodes seen in each destination is
zero and this information is updated each time a node reaches a destination.
Since the domain is continuous, we divided the network area into many small
contiguous cells that represent possible destinations. Each cell has a squared
area, and its size depends on the transmitting range of the nodes. Once a node
reaches a cell, it should be able to communicate with every other node that is
in the same cell at the same time. Hence, the size of the cell is such that its
diagonal is equal to the transmitting radius of the nodes. Based on this, each
node can easily build a \emph{map} of the network area, and can also calculate
the weight for each cell in the map. These information will be used to
determine the next destination: The node chooses its cell
destination randomly and proportionally with its weight, whereas the exact
destination point (remind that the network area is continuous) is taken
uniformly at random over the cell's area. Note that,
according to our experiments, it is not really necessary that the node has a
\emph{full} map of the domain. It can remember just the most popular cells it
has visited and assume that everywhere else there is nobody (until, by chance,
it chooses one of these places as destination and learn that they are indeed
popular). The general properties of SWIM holds as well.

Once a node has chosen its next destination, it starts moving towards it
following a straight line and with a speed that is proportional to the distance
between the starting point and the destination. To keep things simple, in the
simulator the node chooses as its speed value exactly the distance between these
two points. The speed remains constant till the node reaches the destination. In
particular, that means that nodes finish each leg of their movements in constant
time. This can seem quite an oversimplification, however, it is useful and also
not far from reality. Useful to simplify the model; not far from reality since
we are used to move slowly (maybe walking) when the destination is nearby,
faster when it is farther, and extremely fast (maybe by car) when the
destination is far-off.

More specifically, let $A$ be one of the nodes and $h_A$ its home. Let also $C$
be one of the possible destination cells. We will denote with $\textit{seen}(C)$
the number of nodes that node~$A$ encountered in $C$ the last time it reached
$C$. As we already mentioned, this number is $0$ at the beginning of the
simulation and it is updated each time node~$A$ reaches a destination in
cell~$C$. Since $h_A$ is a point, whereas $C$ is a cell, when calculating the
distance of $C$ from its home $h_A$, node~$A$ refers to the center of the cell's
area. In our case, being the cell a square, its center is the mid diagonal
point. The weight that node~$A$ assigns to cell $C$ is as follows:
\begin{equation}
\label{eq:weight}
w(C) = \alpha\cdot\textit{distance}(h_A, C) + (1-\alpha)\cdot\textit{seen}(C).
\end{equation}
where $\textit{distance}(h_A, C)$ is a function that decays as a power law as
the distance between node~$A$ and cell~$C$ increases.

In the above equation $\alpha$ is a constant in $[0;1]$. Since the weight that a
node assigns to a place represents the probability that the node chooses it as
its next destination, the value of $\alpha$ has a strong effect on the node's
decisions---the larger is $\alpha$, the more the node will tend to go to places
near its home. The smaller is $\alpha$, the more the node will tend to go to
``popular'' places. Even if it goes beyond our scope in this paper, we strongly
believe that would be interesting to exploit consequences of using different
values for $\alpha$. We do think that both small and big values for $\alpha$
rise clustering effect of the nodes. In the first case, the clustering effect
is based on the neighborhood locality of the nodes, and is more related to a
social type: Nodes that ``live'' near each other should tend to frequent the
same places, and therefore tend to be ``friends''. In the second case, instead,
the clustering effect should raise as a consequence of the popularity of the
places.

When reaching destination the node decides how long to remain there. One of the
key observations is that in real life a person usually stays for a long time
only in a few places, whereas there are many places where he spends a short
period of time. Therefore, the distribution of the waiting time should follow a
power law. However, this is in contrast with the experimental evidence that
inter-contact time has an exponential cut-off, and with the intuition that, in
many practical scenarios, we won't spend more than a few hours standing at the
same place (our goal is to model day time mobility). So, SWIM uses an upper
bounded power law distribution for waiting time, that is, a truncated power law.
Experimentally, this seems to be the correct choice.

\subsection{Power law and exponential decay dichotomy}

In a recent work~\cite{milan07}, it is observed that the distribution of
inter-contact time in real life experiments shows a so  called dichotomy: First
a power law until a certain point in time, then an exponential cut-off.
In~\cite{cai07mobicom}, the authors suggest that the exponential cut-off is due
to the bounded domain where nodes move. In SWIM, inter-contact time distribution
shows exactly the same dichotomy. More than that, our experiments show that, if
the model is properly tuned, the distribution is strikingly similar to that of
real life experiments.

We show here, with a mathematically rigorous proof, that the distribution of
inter-contact time of nodes in SWIM has an exponential
tail. Later, we will see experimentally that the same distribution has indeed
a head distributed as a power law. Note that the proof has to cope with a
difficulty due to the social nature of SWIM---every decision taken in SWIM by a
node \emph{not} only depends on its own previous decisions, but also depends on
other nodes' decisions: Where a node goes now, strongly affects where it
will choose to go in the future, and, it will affect also where other
nodes will chose to go in the future. So, in SWIM there are no renewal
intervals, decisions influence future decisions of other nodes, and nodes never
``forget'' their past.

In the following, we will consider two nodes $A$ and $B$. Let $A(t)$, $t\ge0$,
be the position of node~$A$ at time~$t$. Similarly, $B(t)$ is the position of
node~$B$ at time~$t$. We assume that at time~$0$ the two nodes are leaving
visibility after meeting. That is, $||A(0)-B(0)||=r$, $||A(t)-B(t)||<r$ for
$t\in 0^-$, and $||A(t)-B(t)||>r$ for $t\in 0^+$. Here, $||\cdot||$ is the
euclidean distance in the square. The inter-contact time of nodes $A$ and $B$
is defined as:
\begin{equation*}
T_I=\inf_{t>0} \{t:||A(t)-B(t)||\le r\}
\end{equation*}
\begin{assumption}
\label{ass:lower}
For all nodes~$A$ and for all cells~$C$, the distance function $distance(A,C)$
returns at least $\mu>0$.
\end{assumption}

\begin{theorem}
If $\alpha>0$ and under Assumption~\ref{ass:lower}, \emph{the tail} of the
inter-contact time distribution between nodes~$A$ and~$B$ in SWIM has an
exponential decay.
\end{theorem}
\begin{IEEEproof}
To prove the presence of the exponential cut-off, we will show that there exists
constant $c>0$ such that
\begin{equation*}
\mathbb{P}\{T_I>t\}\le e^{-ct}
\end{equation*}
for all sufficiently large $t$. Let $t_i=i\lambda$, $i=1,2,\dotsc$, be
a sequence of times. Constant $\lambda$ is large enough that each node has
to make a way point decision in the interval between $t_i$ and $t_{i+1}$
and that each node has enough time to finish a leg. Recall that this is of
course possible since waiting time at way points is bounded above and since
nodes complete each leg of movement in constant time. The idea is to take
snapshots of nodes $A$ and $B$ and see whether they see each other at each
snapshot. However, in the following, we also need that at least one of the two
nodes is not moving at each snapshot. So, let
\begin{equation*}
\begin{split}
\delta_i=\text{min}\{ & \delta\ge 0 : \text{either $A$ or $B$ is}\\
& \text{at a way point at time $t_i+\delta$}\}.
\end{split}
\end{equation*}
Clearly, $t_i+\delta_i<t_{i+1}$, for all $i=1,2,\dotsc$.

We take the sequence of snapshots $\{t_i+\delta_i\}_{i>0}$. Let $\epsilon_i=\{||A(t_i+\delta_i)-B(t_i+\delta_i)||>r\}$ be the event that nodes $A$ and $B$ are not in visibility range at time $t_i+\delta_i$. We have that
\begin{equation*}
\mathbb{P}\{T_I>t\}\le \mathbb{P}\left\{\bigcap_{i=1}^{\lfloor t/\lambda\rfloor
-1}
\epsilon_i\right\}=\prod_{i=1}^{\lfloor t/\lambda\rfloor -1}
\mathbb{P}\{ \epsilon_i| \epsilon_{i-1}\cdots\epsilon_1\}.
\end{equation*}
Consider $\mathbb{P}\{ \epsilon_i| \epsilon_{i-1}\cdots\epsilon_1\}$. At
time~$t_i+\delta_i$, at least one of the two nodes is at a way point, by
definition of $\delta_i$. Say node~$A$, without loss of generality. Assume that
node~$B$ is in cell $C$ (either moving or at a way point). During its last way
point decision, node~$A$ has chosen cell $C$ as its next way point with
probability at least $\alpha\mu>0$, thanks to Assumption~\ref{ass:lower}. If
this is the case, the two nodes~$A$ and~$B$ are now in visibility. Note that the
decision has been made after the previous snapshot, and that it is not
independent of previous decisions taken by node~$A$, and it is not even
independent of previous decisions taken by node~$B$ (since the social nature of
decisions in SWIM). Nonetheless, with probability at least $\alpha\mu$ the two
nodes are now in visibility. Therefore,
\begin{equation*}
\mathbb{P}\{ \epsilon_i| \epsilon_{i-1}\cdots\epsilon_1\}\le 1-\alpha\mu.
\end{equation*}
So,
\begin{equation*}
\begin{split}
\mathbb{P}\{T_I>t\} & \le \mathbb{P}\left\{\bigcap_{i=1}^{\lfloor
t/\lambda\rfloor -1}
\epsilon_i\right\}=\prod_{i=1}^{\lfloor t/\lambda\rfloor -1}
\mathbb{P}\{ \epsilon_i| \epsilon_{i-1}\cdots\epsilon_1\}\\
 & \le (1-\alpha\mu)^{\lfloor t/\lambda\rfloor -1}\sim e^{-ct},
\end{split}
\end{equation*}
for sufficiently large $t$.
\end{IEEEproof}

\section{Real traces}
\begin{table*}
\begin{center}
\begin{tabular}{|c|c|c|c|c|c|c|}
\hline
Experimental data set \T \B & Cambridge~05 & Cambridge~06 & Infocom~05\\ \hline
Device \T & iMote & iMote & iMote\\
Network type & Bluetooth & Bluetooth & Bluetooth\\
Duration (days)& 5 & 11 & 3\\
Granularity (sec)& 120 & 600 & 120\\
Devices number & 12 & 54 (36 mobile) & 41\\
Internal contacts number& 4,229 & 10,873 & 22,459\\
Average Contacts/pair/day & 6.4 & 0.345 & 4.6\\[1mm] \hline
\end{tabular}
\caption{The three experimental data sets}
\label{tab:realtraces}
\end{center}
\end{table*}

In order to show the accuracy of SWIM in simulating real life scenarios, we will
compare SWIM with three traces gathered during experiments done with real
devices carried by people. We will refer to these traces as \emph{Infocom~05},
\emph{Cambridge~05} and \emph{Cambridge~06}. Characteristics of these data sets
such as inter-contact and contact distribution have been observed in several
previous works~\cite{hui05, leguay06,hui06}.
\begin{itemize}
\item In \emph{Cambridge 05}~\cite{cambridge05} the authors used Intel iMotes to
collect the data. The iMotes were distributed to students of the University of
Cambridge and were programmed to log contacts of all visible mobile devices. The
number of devices that were used for this experiment is 12. This data set covers
5 days.
\item In \emph{Cambridge 06}~\cite{upmcCambridgeData} the authors repeated the
experiment using more devices. Also, a number of stationary nodes were deployed
in various locations around the city of Cambridge UK. The data of the stationary
iMotes will not be used in this paper. The number of mobile devices used is 36
(plus 18 stationary devices). This data set covers 11 days.
\item In \emph{Infocom~05} ~\cite{cambridgeInfocomData} the same devices as in
\emph{Cambridge} were distributed to students attending the Infocom 2005 student
workshop. The number of devices is 41. This experiment covers approximately 3
days.
\end{itemize}
Further details on the real traces we use in this paper are shown in
Table~\ref{tab:realtraces}.

\section{SWIM vs Real traces}
\label{sec:experiments}

\subsection{The simulation environment}
In order to evaluate SWIM, we built a discrete even simulator of the model.
The simulator takes as input
\begin{itemize}
 \item $n$: the number of nodes in the network;
  \item $r$: the transmitting radius of the nodes;
  \item the simulation time in seconds;
  \item coefficient $\alpha$ that appears in Equation~\ref{eq:weight};
  \item the distribution of the waiting time at destination.
\end{itemize}
The output of the simulator is a text file containing records on each main event
occurrence. The main events of the system and the related outputs are:
\begin{itemize}
\item \emph{Meet} event: When two nodes are in range with each other. The
output line contains the ids of the two nodes involved and the time of
occurrence.
\item \emph{Depart} event: When two nodes that were in range of each other are
not anymore. The output line contains the ids of the two nodes involved and the
time of occurrence.
\item \emph{Start} event: When a node leaves its current location and starts
moving towards destination. The output line contains the id of the location, the
id of the node and the time of occurrence.
\item \emph{Finish} event: When a node reaches its destination. The output line
contains the id of the destination, the id of the node and the time of
occurrence.
\end{itemize}
In the output, we don't really need information on the geographical position of
the nodes when the event occurs. However, it is just straightforward to extend
the format of the output file to include this information. In this form, the
output file contains enough information to compute correctly inter-contact
intervals, number of contacts, duration of contacts, and to implement state of
the art forwarding protocols.

During the simulation, the simulator keeps a vector $\textit{seen}(C)$ updated
for each sensor. Note that the nodes do not necessarily agree on what is the
popularity of each cell. As mentioned earlier, it is not necessary to keep in
memory the whole vector, without changing the qualitative behavior of the mobile
system. However, the three scenarios Infocom~05, Cambridge~05, and Cambridge~06
are not large enough to cause any real memory problem. Vector~$\textit{seen}(C)$
is updated at each \emph{Finish} and \emph{Start} event, and is not changed
during movements.

\subsection{The experimental results}

In this section we will present some experimental results in order to show that
SWIM is a simple and good way to generate synthetic traces with the same
statistical properties of real life mobile scenarios.
\begin{figure}[!ht]
\centering
\subfigure[Distribution of the inter-contact time in Infocom~05 and in SWIM]{
\centering
\includegraphics[width=.4\textwidth]{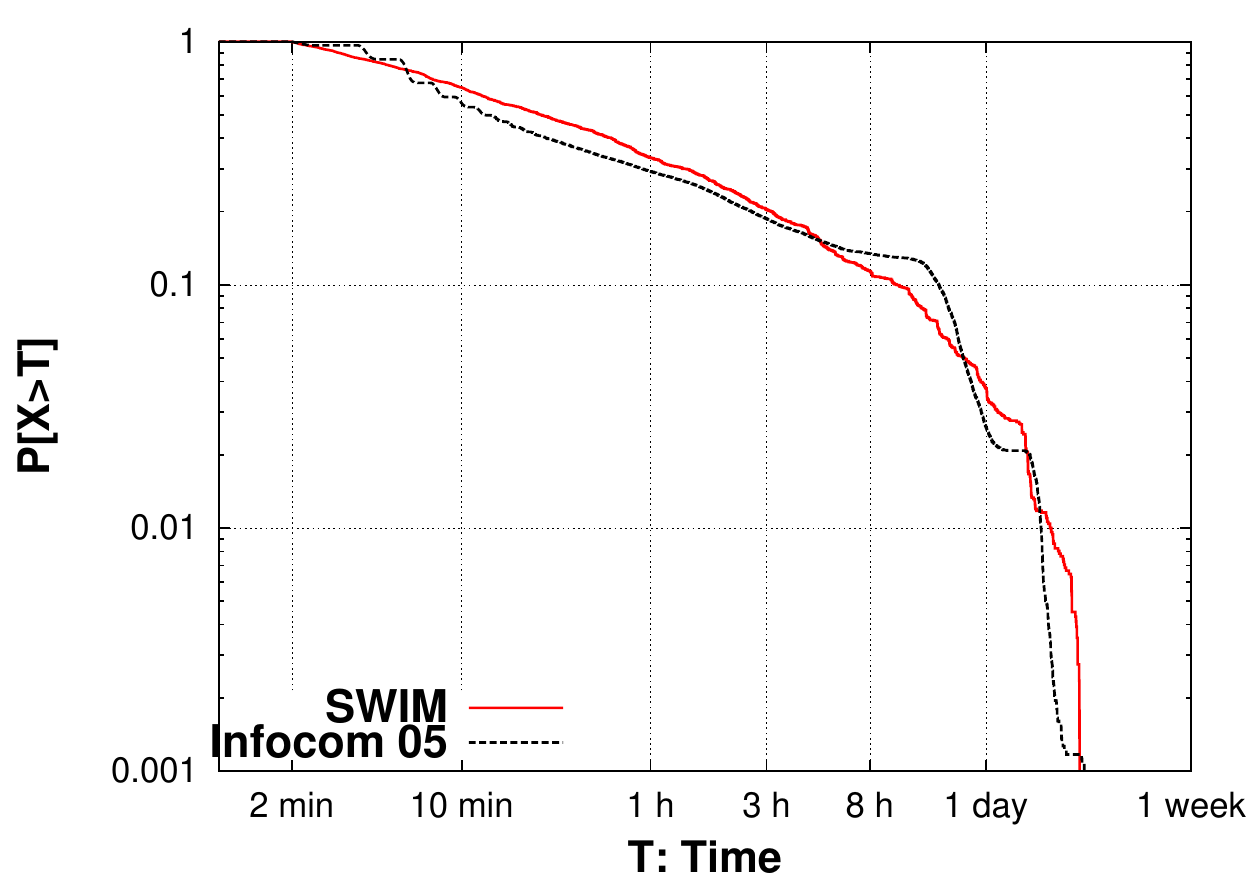}
\label{fig:ICT infocom}}
\qquad
\subfigure[Distribution of the contact duration for each pair of nodes in
Infocom~05 and in SWIM]{
\centering
\includegraphics[width=.4\textwidth]{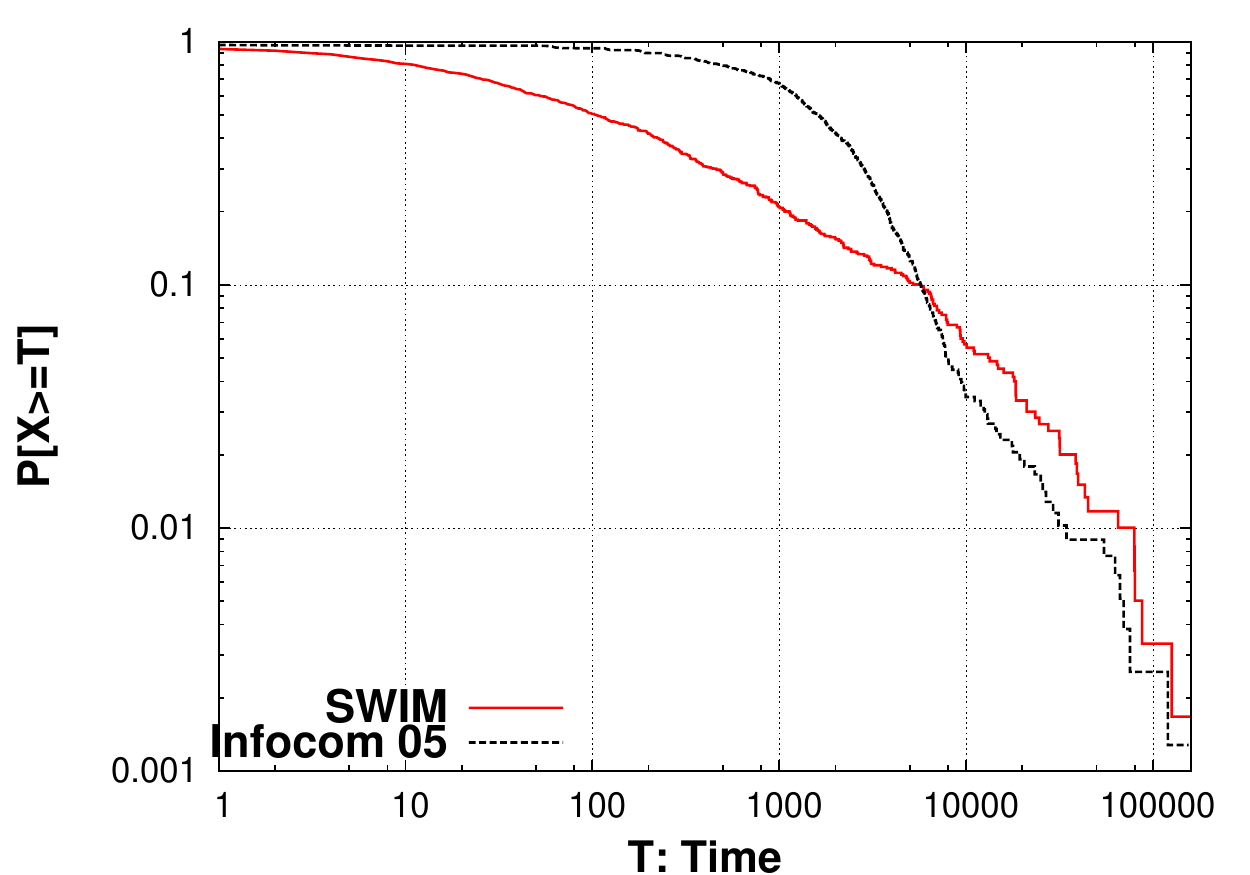}
\label{fig:CONT infocom}}
\qquad
\subfigure[Distribution of the number of contacts for each pair of nodes in
Infocom~05 and in SWIM]{
\centering
\includegraphics[width=.4\textwidth]{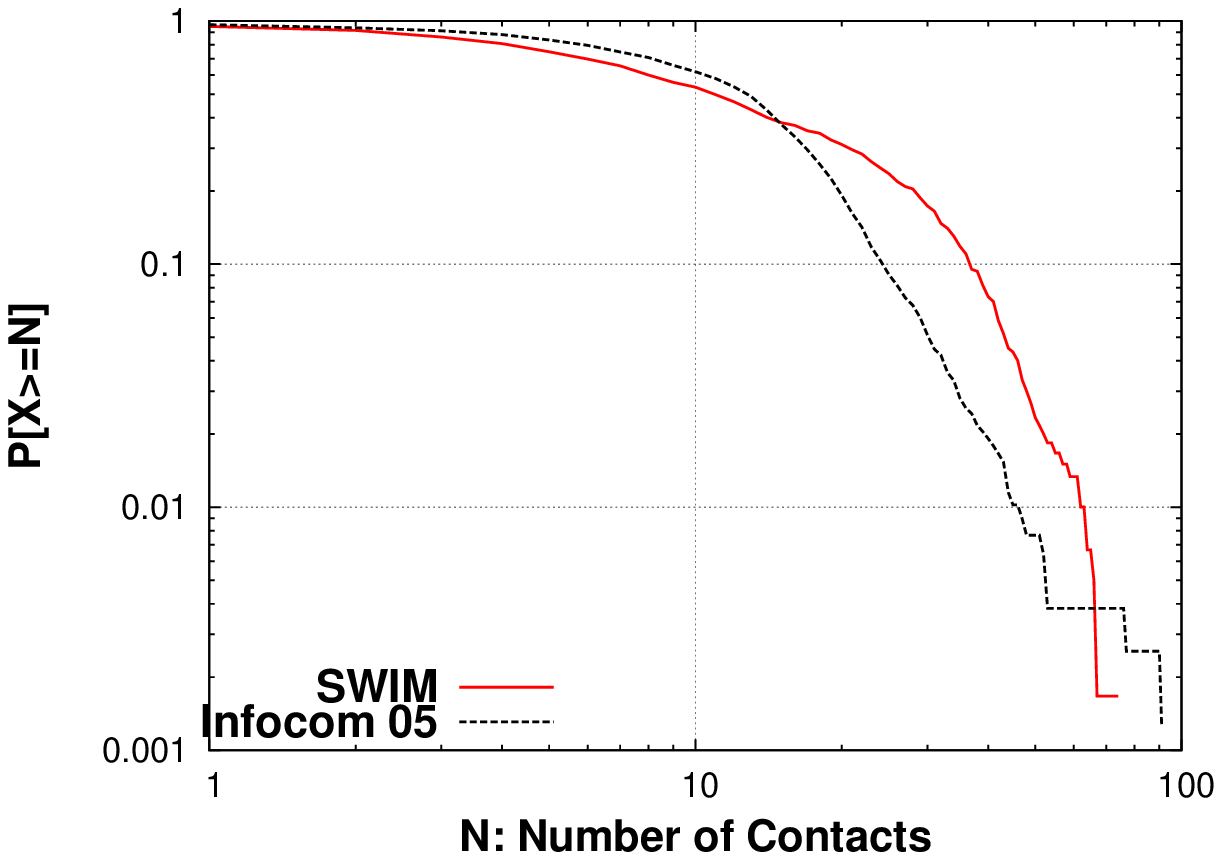}
\label{fig:CONT-NR infocom}}
\caption{SWIM and Infocom~05}
\label{fig:infocom}
\end{figure}
\begin{figure}[t]
\centering
\subfigure[Distribution of the inter-contact time in Cambridge~05 and in SWIM]{
\centering
\includegraphics[width=.4\textwidth]{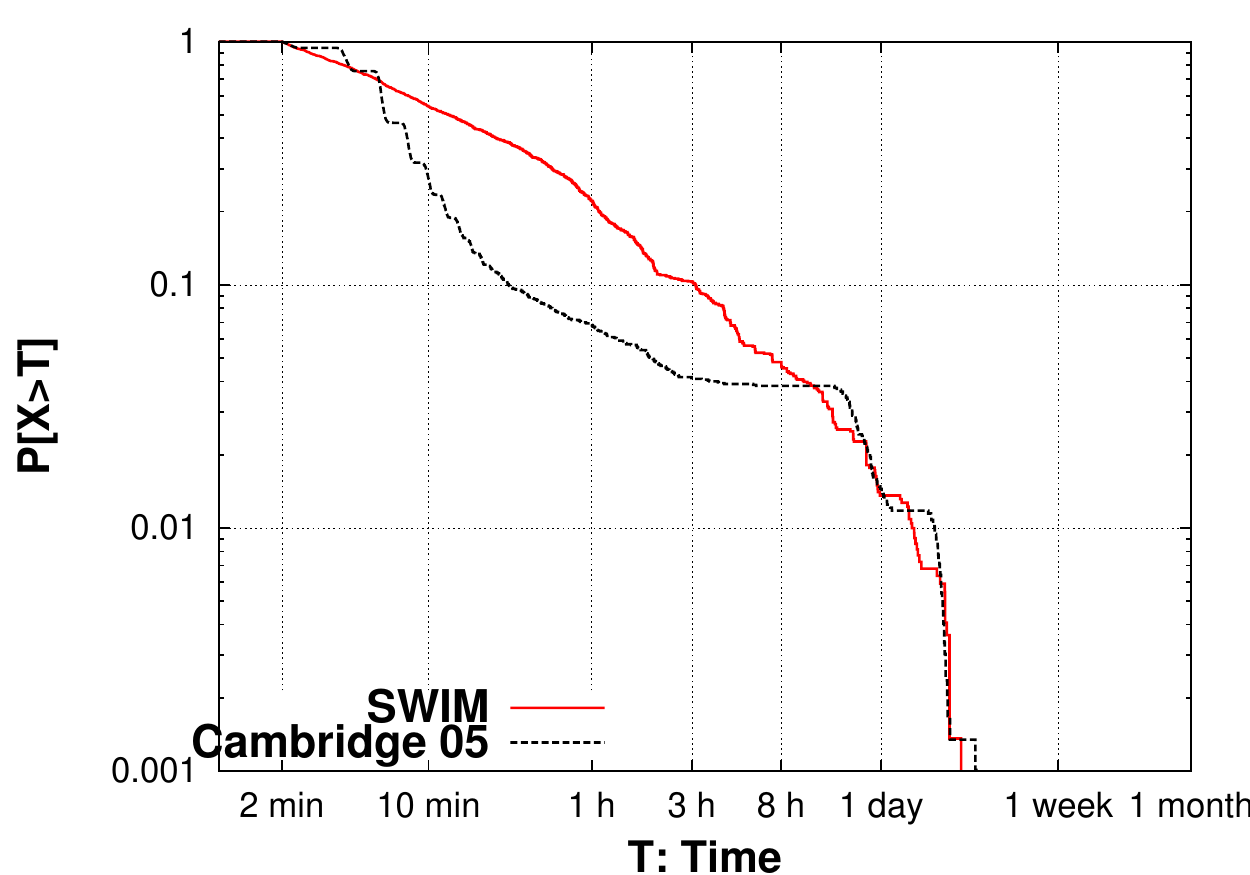}
\label{fig:ICT cambridge05}}
\qquad
\subfigure[Distribution of the contact duration for each pair of nodes in
Cambridge~05 and in SWIM]{
\centering
\includegraphics[width=.4\textwidth]{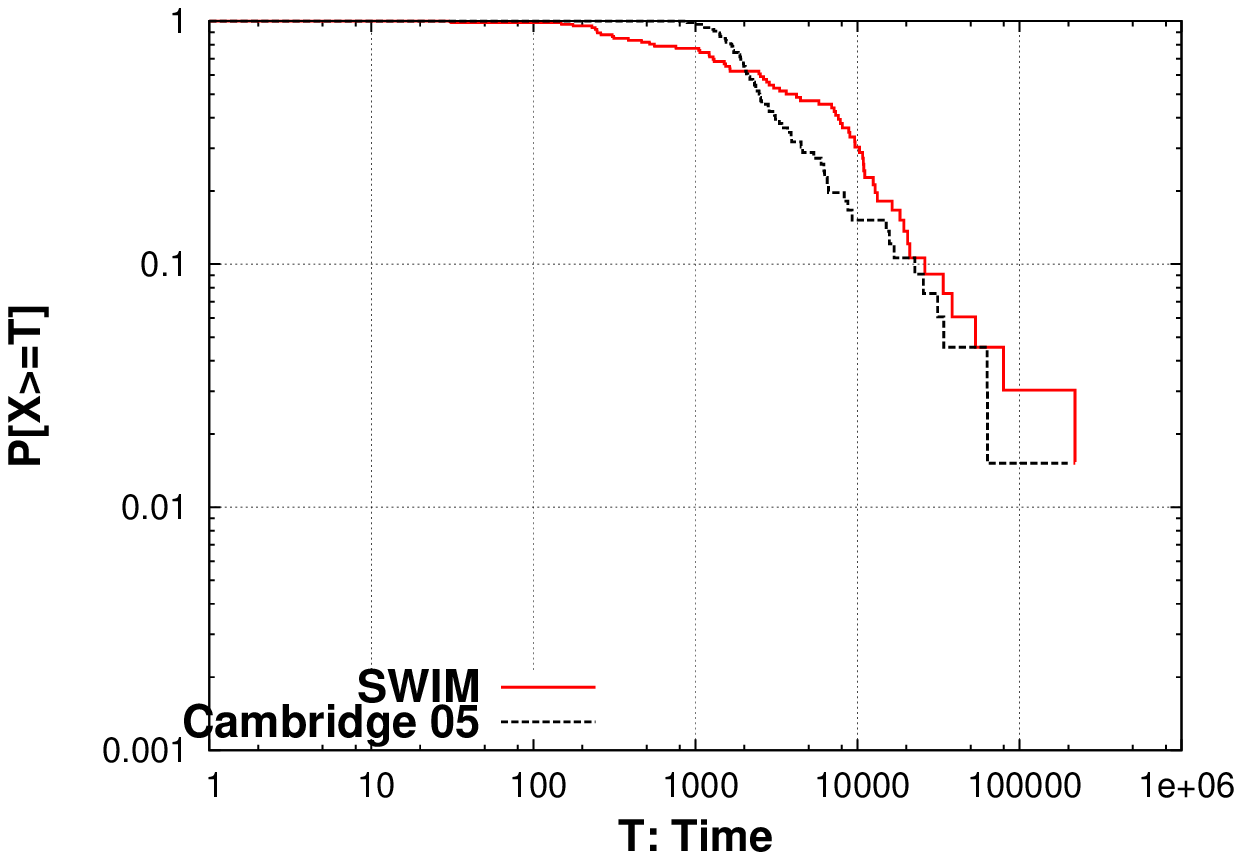}
\label{fig:CONT cambridge05}}
\qquad
\subfigure[Distribution of the number of contacts for each pair of nodes in
Cambridge~05 and in SWIM]{
\centering
\includegraphics[width=.4\textwidth]{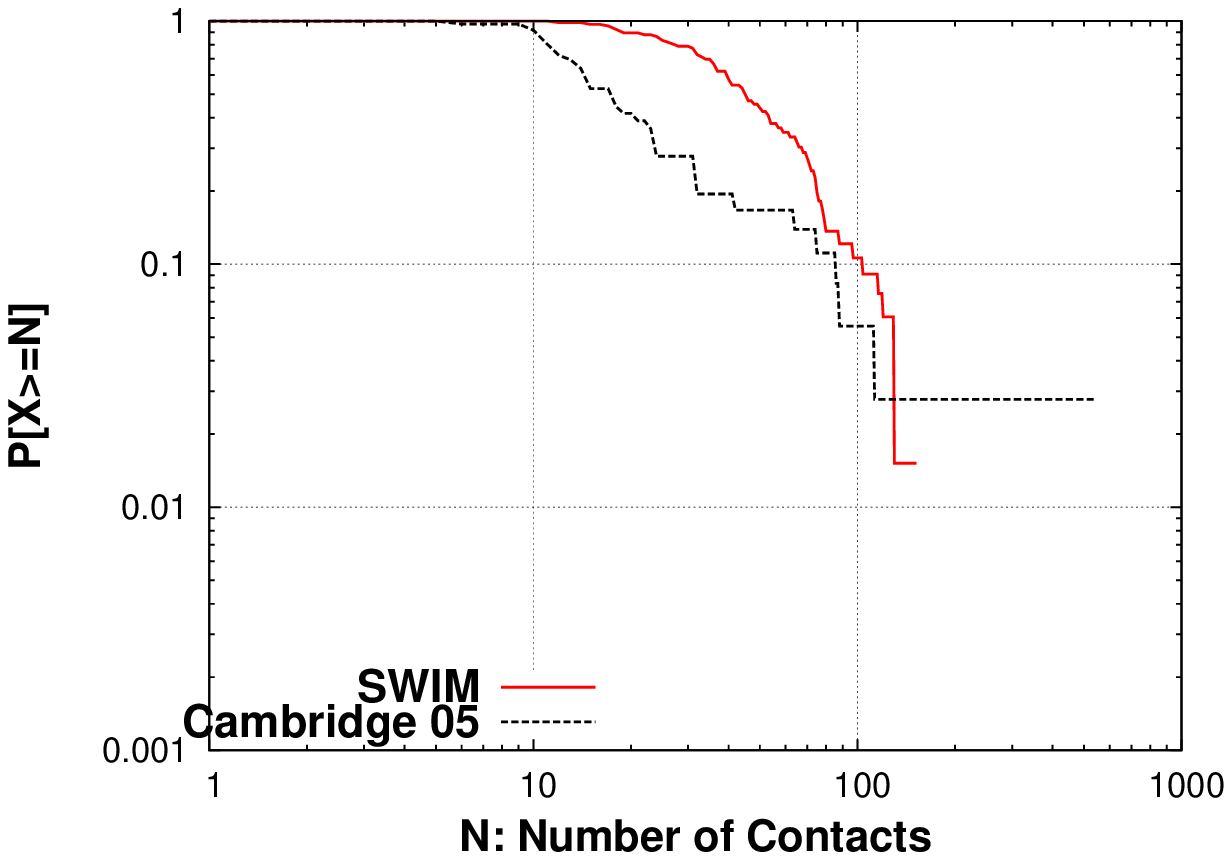}
\label{fig:CONT-NR cambridge05}}
\caption{SWIM and Cambridge~05}
\label{fig:cambridge05}
\end{figure}
\begin{figure}[t]
\centering
\subfigure[Distribution of the inter-contact time in Cambridge~06 and in SWIM]{
\centering
\includegraphics[width=.4\textwidth]{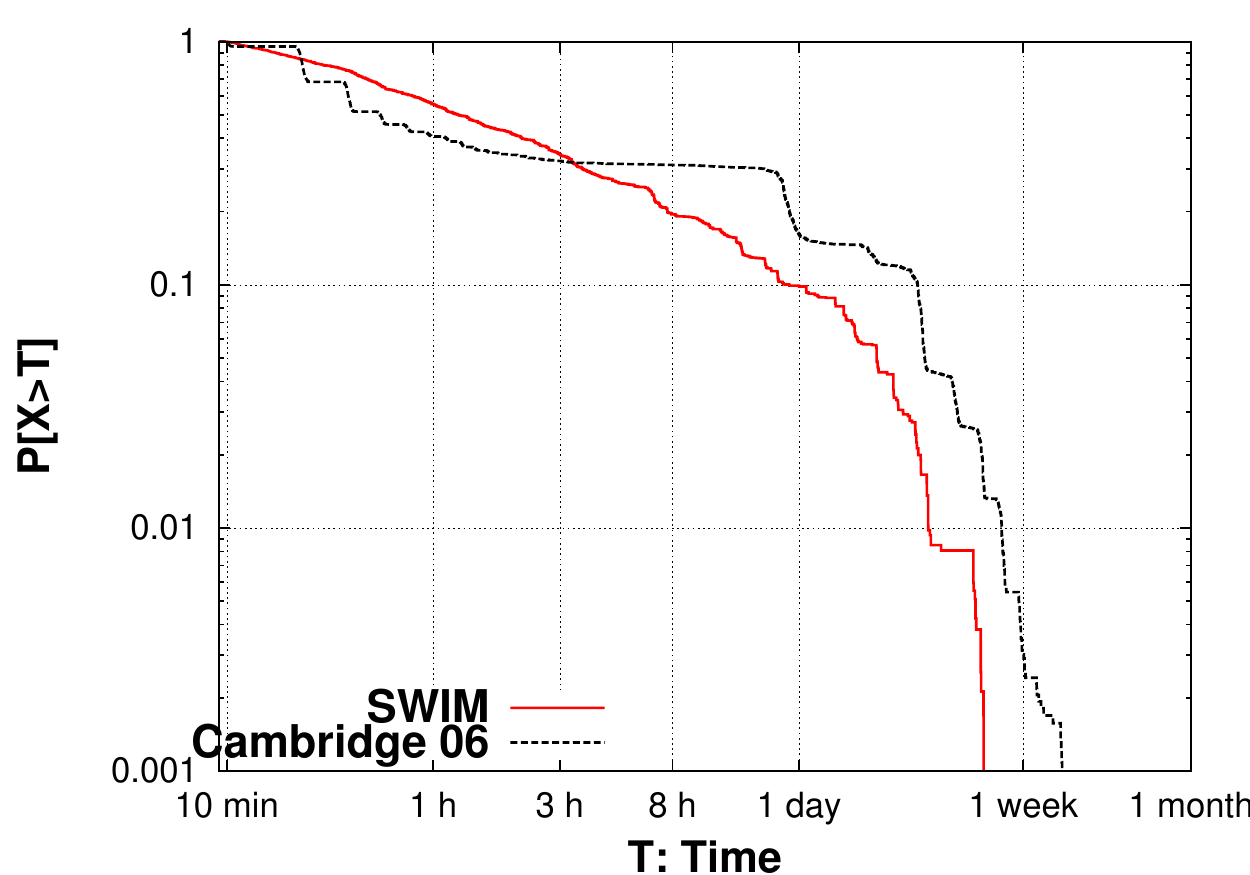}
\label{fig:ICT cambridge}}
\qquad
\subfigure[Distribution of the contact duration for each pair of nodes in
Cambridge~06 and in SWIM]{
\centering
\includegraphics[width=.4\textwidth]{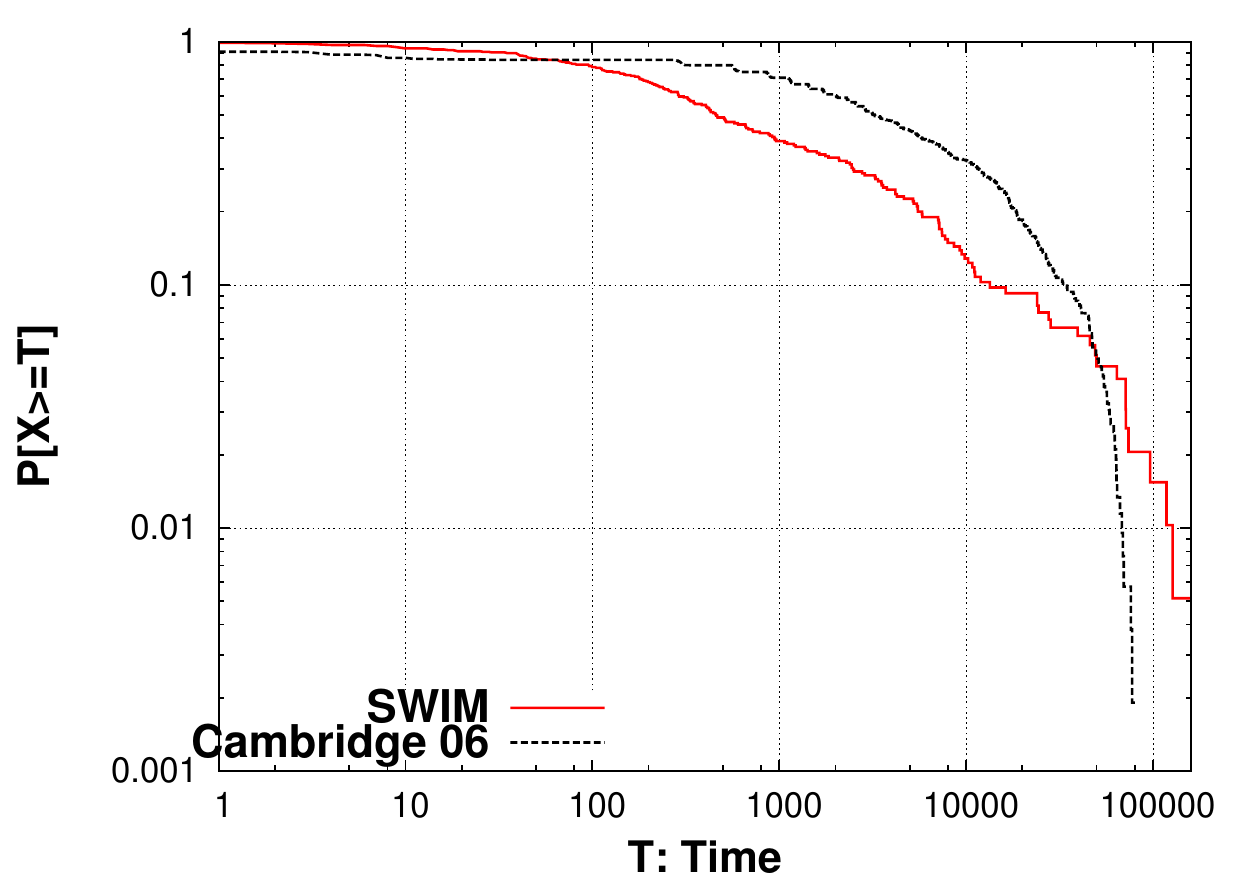}
\label{fig:CONT cambridge}}
\qquad
\subfigure[Distribution of the number of contacts for each pair of nodes in
Cambridge~06 and in SWIM]{
\centering
\includegraphics[width=.4\textwidth]{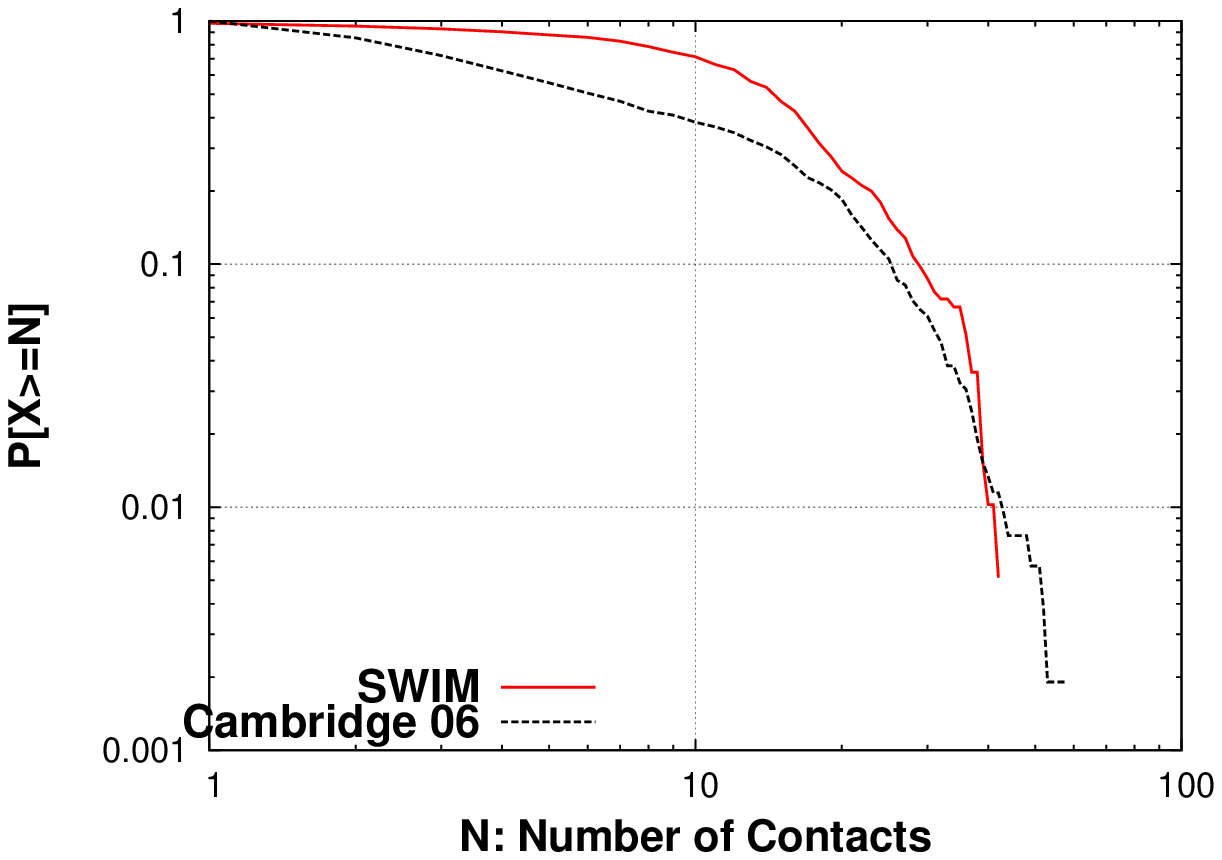}
\label{fig:CONT-NR cambridge}}
\caption{SWIM and Cambridge~06}
\label{fig:cambridge06}
\end{figure}
The idea is to tune the few parameters used by SWIM in order to simulate
Infocom~05, Cambridge~05, and Cambridge~06. For each of the experiments we
consider the following metrics: inter-contact time CCD function, contact
distribution per pair of nodes, and number of contacts per pair of nodes. The
inter-contact time distribution is important in mobile networking since it
characterizes the frequency with which information can be transferred between
people in real life. It has been widely studied for real traces in a large
number of previous papers~\cite{hui05, hui06, leguay06, cai07mobicom,
milan07, musolesi07, cai08mobihoc}. The contact distribution per pair
of nodes and the number of contacts per pair of nodes are also important. Indeed
they represent a way to measure relationship between people. As it was also
discussed in~\cite{hui07community, hui07socio, hui08mobihoc} it's natural to
think that if a couple of people spend more time together and meet each other
frequently they are familiar to each other. Familiarity is important in
detecting communities, which may help improve significantly the design and
performance of forwarding protocols in mobile environments such as
DNTs~\cite{hui08mobihoc}. Let's now present the experimental results obtained
with SWIM when simulating each of the real scenarios of data sets.
Since the scenarios we consider use iMotes, we model our network node according
to iMotes properties (outdoor range $30\textrm{m}$). We initially distribute the
nodes over a network area of size $300\times300~\textrm{m}^2$. In the following,
we assume for the sake of simplicity that the network area is a square of side
1, and that the node transmission range is 0.1. In all the three experiments we
use a power law with slope $a=1.45$ in order to generate waiting time values of
nodes when arriving to destination, with an upper bound of 4 hours. We use as
$\textit{seen}(C)$ function the fraction of the nodes seen in cell~$C$, and as
$\textit{distance}(x,C)$ the following
\begin{equation*}
\textit{distance}(x,C)=\frac{1}{\left(1+k||x-y||\right)^2},
\end{equation*}
where $x$ is the position of the home of the current node, and $y$ is the
position of the center of cell~$C$. Positions are coordinates in the square of
size 1. Constant $k$ is a scaling factor, set to $0.05$, which accounts for the
small size of the experiment area. Note that function $\textit{distance}(x,C)$
decays as a power law. We come up with this choice after a large set of
experiments, and the choice is heavily influenced by scaling factors.

We start with Infocom~05. The number of nodes $n$ and the simulation time are
the same as in the real data set, hence 41 and 3 days respectively. Since the
area of the real experiment was quite small (a large hotel), we deem that
$300\times300~m^2$ can be a good approximation of the real scenario. In
Infocom~05, there were many parallel sessions. Typically, in such a case one
chooses to follow what is more interesting to him. Hence, people with the same
interests are more likely to meet each other. In this experiment, the parameter
$\alpha$ such that the output fit best the real traces is $\alpha=0.75$. The
results of this experiment are shown in Figure~\ref{fig:infocom}.

We continue with the Cambridge scenario. The number of nodes and the simulation
time are the same as in the real data set, hence 11 and 5 days respectively. In
the Cambridge data set, the iMotes were distributed to two groups of students,
mainly undergrad year~1 and~ 2, and also to some PhD and Master students.
Obviously, students of the same year are more likely to see each other more
often. In this case, the parameter $\alpha$ which best fits the real traces is
$\alpha=0.95$. This choice proves to be fine for both Cambridge~05 and
Cambridge~06. The results of this experiment are shown in
Figure~\ref{fig:cambridge05} and~\ref{fig:cambridge06}.

In all of the three experiments, SWIM proves to be an excellent way to generate
synthetic traces that approximate real traces. It is particularly interesting
that the same choice of parameters gets goods results for all the metrics under
consideration at the same time.

\section{Comparative performance of forwarding protocols}
\label{sec:forwarding}
\begin{figure*}
\label{fig:forwarding}
\centering
\subfigure{
\centering
\includegraphics[width=.31\textwidth]{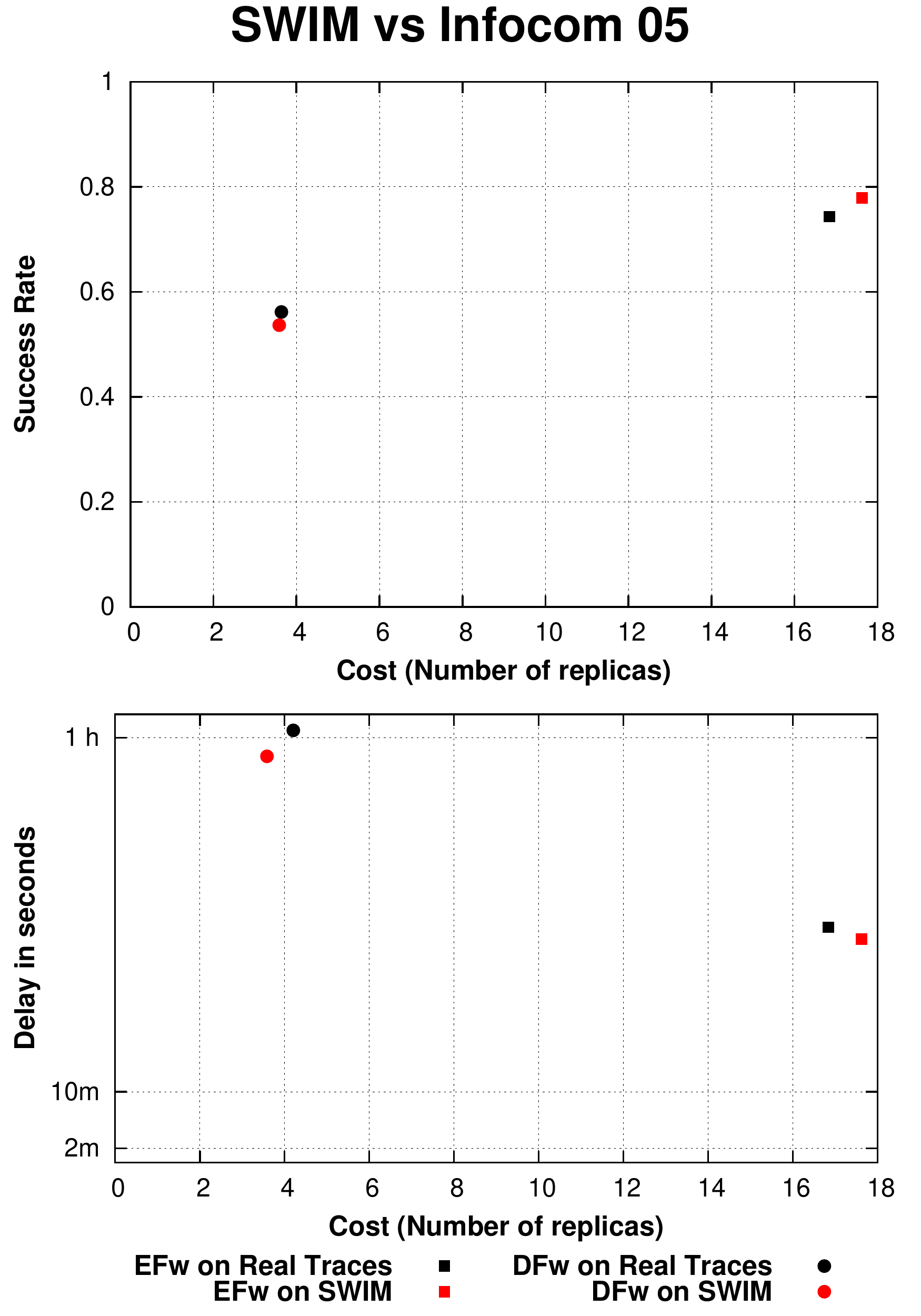}
\label{fig:perf infocom}}
\subfigure{
\centering
\includegraphics[width=.31\textwidth]{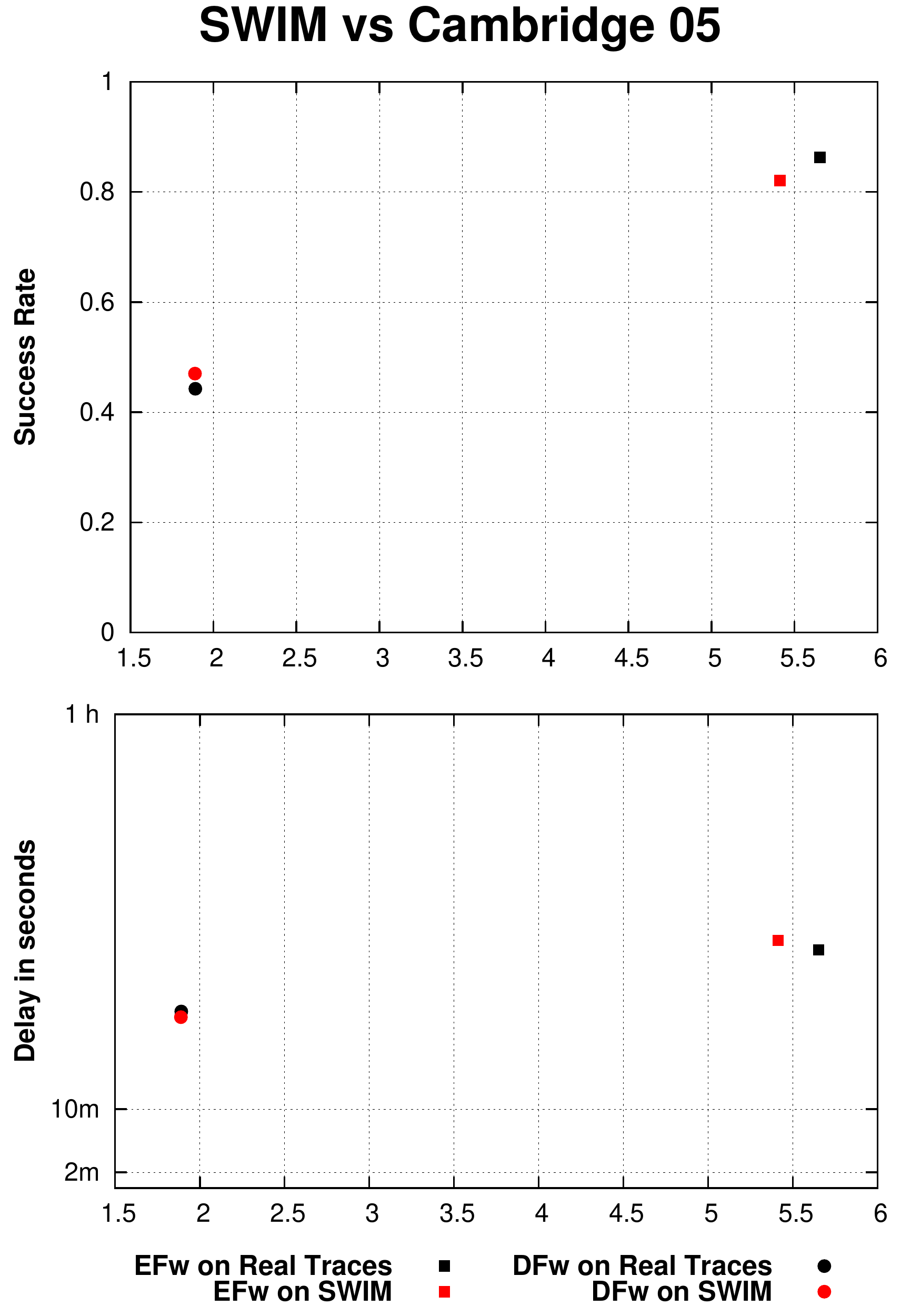}
\label{fig:perf cambridge05}}
\subfigure{
\centering
\includegraphics[width=.31\textwidth]{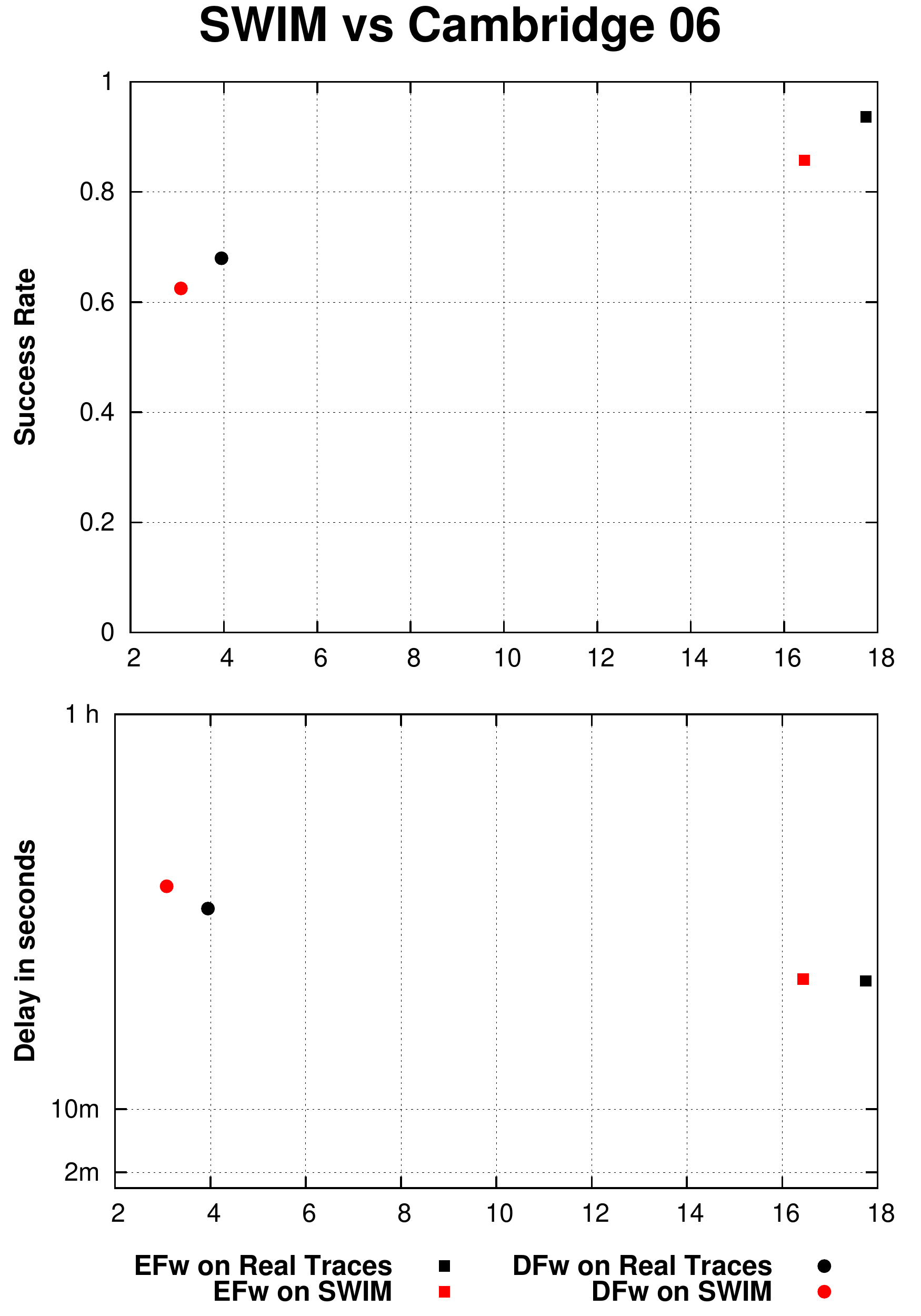}
\label{fig:perf cambridge 06}}
\caption{Performance of both forwarding protocols on real traces and SWIM
traces. EFw denotes Epidemic Forwarding while DFwd Delegation Forwarding.}
\label{fig:performance}
\end{figure*}

In this section we show other experimental results of SWIM, related to
evaluation of two simple forwarding protocols for DNTs such as Epidemic
Forwarding~\cite{vahdat00epidemic} and simplified version of Delegation
Forwarding\cite{dfw08} in which each node has a random constant as its quality.
Of course, this simplified version of delegation forwarding is not very
interesting and surely non particularly efficient. However, we use it just as a
worst case benchmark against epidemic forwarding, with the understanding that
our goal is just to validate the quality of SWIM, and not the quality of the
forwarding protocol.

In the following experiments, we use for each experiment the same tuning used
in the previous section. That is, the parameters input to SWIM are not
``optimized'' for each of the forwarding protocols, they are just the same that
has been used to fit real traces with synthetic traces.

For the evaluation of the two forwarding protocols we use the same assumptions
and the same way of generating traffic to be routed as in~\cite{dfw08}.
For each trace and forwarding protocol a set of messages is generated with
sources and destinations chosen uniformly at random, and generation times form a
Poisson process averaging one message per 4 seconds.
The nodes are assumed to have infinite buffers and carry all message replicas
they receive until the end of the simulation. The metrics we are concerned with
are: \emph{cost}, which is the number of replicas per generated message;
\emph{success rate} which is the fraction of generated messages for which at
least one replica is delivered; \emph{average delay} which is the average
duration per delivered message from its generation time to the first arrival of
one of its replicas.
As in \cite{dfw08} we isolated 3-hour periods for each data trace (real and
synthetic) for our study. Each simulation runs therefore 3 hours. to avoid
end-effects no messages were generated in the last hour of each trace.

In the two forwarding protocols, upon contact with node $A$, node $B$ decides
which message from its message queue to forward in the following way:
\begin{trivlist}
\item
\textbf{Epidemic Forwarding:} Node $A$ forwards message~$m$ to node $B$ unless
$B$ already has a replica of $m$. This protocol achieves the best possible
performance, so it yields upper bounds on success rate and average delay.
However, it does also have a high cost.
\item
\textbf{(Simplified) Delegation Forwarding:} To each node is initially given a
quality, distributed uniformly in $(0;1]$. To each message is given a rate,
which, in every instant corresponds to the quality of the node with the best
quality that message have seen so far. When generated the message inherits the
rate from the node that generates it (that would be the sender for that
message). Node $A$ forwards message $m$ to node $B$ if the quality of node $B$
is greater than the rate of the copy of $m$ that $A$ holds. If $m$ is forwarded
to $B$, both nodes $A$ and $B$ update the rate of their copy of $m$ to the
quality of $B$.
\end{trivlist}

Figure~\ref{fig:forwarding} shows how the two forwarding protocols perform in
both real and synthetic traces, generated with SWIM.
As you can see, the results are excellent---SWIM predicts very accurately the
performance of both protocols. Most importantly, this is not due to a customized
tuning that has been optimized for these forwarding protocols, it is just the
same output that SWIM has generated with the tuning of the previous section.
This can be important methodologically: To tune SWIM on a particular scenario,
you can concentrate on a few well known and important statistical properties
like inter-contact time, number of contacts, and duration of contacts. Then, you
can have a good confidence that the model is properly tuned and usable to get
meaningful estimation of the performance of a forwarding protocol.

\section{Conclusions}
\label{sec:conclusions}
In this paper we present SWIM, a new mobility model for ad hoc networking. SWIM
is simple, proves to generate traces that look real, and provides an accurate
estimation of forwarding protocols in real mobile networks. SWIM can be used to
improve our understanding of human mobility, and it can support theoretical work
and it can be very useful to evaluate the performance of networking protocols in
scenarios that scales up to very large mobile systems, for which we don't have
real traces.
\IEEEtriggeratref{7}
\bibliographystyle{ieeetr}


\begin{thebibliography}{10}

\bibitem{camp02wcmc}
T.~Camp, J.~Boleng, and V.~Davies, ``A survey of mobility models for ad hoc
  network research,'' {\em Wireless Communications and Mobile Computing Special
  issue on Mobile Ad Hoc Networking: Research, Trends and Applications},
  vol.~2, no.~5, pp.~483--502, 2002.

\bibitem{rwp}
D.~B. Johnson and D.~A. Maltz, ``Dynamic source routing in ad hoc wireless
  networks,'' in {\em Mobile Computing} (Imielinski and Korth, eds.), vol.~353,
  Kluwer Academic Publishers, 1996.

\bibitem{toronto}
J.~Su, A.~Chin, A.~Popivanova, A.~Goel, and E.~de~Lara, ``User mobility for
  opportunistic ad-hoc networking,'' in {\em WMCSA '04: Proceedings of the
  Sixth IEEE Workshop on Mobile Computing Systems and Applications},
  (Washington, DC, USA), pp.~41--50, IEEE Computer Society, 2004.

\bibitem{hui05}
P.~Hui, A.~Chaintreau, J.~Scott, R.~Gass, J.~Crowcroft, and C.~Diot, ``Pocket
  switched networks and human mobility in conference environments,'' in {\em
  WDTN '05: Proceeding of the 2005 ACM SIGCOMM workshop on Delay-tolerant
  networking}, pp.~244--251, ACM Press, 2005.

\bibitem{hui06}
A.~Chaintreau, P.~Hui, J.~Crowcroft, C.~Diot, R.~Gass, and J.~Scott, ``Impact
  of human mobility on the design of opportunistic forwarding algorithms,'' in
  {\em INFOCOM 2006. 25th IEEE International Conference on Computer
  Communications. Proceedings}, 2006.

\bibitem{milan07}
T.~Karagiannis, J.-Y.~L. Boudec, and M.~Vojnovi\'{c}, ``Power law and
  exponential decay of inter contact times between mobile devices,'' in {\em
  MobiCom '07: Proceedings of the 13th annual ACM international conference on
  Mobile computing and networking}, pp.~183--194, ACM, 2007.

\bibitem{UCAM-CL-TR-617}
A.~Chaintreau, P.~Hui, J.~Crowcroft, C.~Diot, R.~Gass, and J.~Scott, ``Pocket
  switched networks: Real-world mobility and its consequences for opportunistic
  forwarding,'' tech. rep., Computer Laboratory, University of Cambridge, 2006.

\bibitem{cai07mobicom}
H.~Cai and D.~Y. Eun, ``Crossing over the bounded domain: from exponential to
  power-law inter-meeting time in manet,'' in {\em MobiCom '07: Proceedings of
  the 13th annual ACM international conference on Mobile computing and
  networking}, pp.~159--170, ACM, 2007.

\bibitem{vahdat00epidemic}
A.~Vahdat and D.~Becker, ``Epidemic routing for partially connected ad hoc
  networks,'' Tech. Rep. CS-200006, Duke University, 2000.

\bibitem{dfw08}
V.~Erramilli, M.~Crovella, A.~Chaintreau, and C.~Diot, ``Delegation
  forwarding,'' in {\em MobiHoc '08: Proceedings of the 9th ACM international
  symposium on Mobile ad hoc networking and computing}, pp.~251--260, ACM,
  2008.

\bibitem{levy}
I.~Rhee, M.~Shin, S.~Hong, K.~Lee, and S.~Chong, ``On the levy-walk nature of
  human mobility,'' in {\em INFOCOM 2008. IEEE International Conference on
  Computer Communications. Proceedings}, 2008.

\bibitem{musolesi07}
M.~Musolesi and C.~Mascolo, ``Designing mobility models based on social network
  theory,'' {\em SIGMOBILE Mob. Comput. Commun. Rev.}, vol.~11, no.~3,
  pp.~59--70, 2007.

\bibitem{LCA-CONF-2008-049}
M.~Piorkowski, N.~Sarafijanovic-Djukic, and M.~Grossglauser, ``On {C}lustering
  {P}henomenon in {M}obile {P}artitioned {N}etworks,'' in {\em The {F}irst
  {ACM} {SIGMOBILE} {I}nternational {W}orkshop on {M}obility {M}odels for
  {N}etworking {R}esearch}, ACM, 2008.

\bibitem{workingDay}
F.~Ekman, A.~Keränen, J.~Karvo, and J.~Ott, ``Working day movement model,'' in
  {\em The {F}irst {ACM} {SIGMOBILE} {I}nternational {W}orkshop on {M}obility
  {M}odels for {N}etworking {R}esearch}, ACM, 2008.

\bibitem{barabasi08}
M.~C. Gonzalez, C.~A. Hidalgo, and A.-L. Barabasi, ``Understanding individual
  human mobility patterns,'' {\em Nature}, vol.~453, pp.~779--782, june 2008.

\bibitem{leguay06}
J.~Leguay, A.~Lindgren, J.~Scott, T.~Friedman, and J.~Crowcroft,
  ``Opportunistic content distribution in an urban setting,'' in {\em CHANTS
  '06: Proceedings of the 2006 SIGCOMM workshop on Challenged networks},
  pp.~205--212, ACM, 2006.

\bibitem{cambridge05}
J.~Scott, R.~Gass, J.~Crowcroft, P.~Hui, C.~Diot, and A.~Chaintreau,
  ``{CRAWDAD} trace cambridge/haggle/imote/cambridge (v. 2006--01--31).''
  Downloaded from
  http://craw\-dad.cs.dart\-mouth.edu/cam\-bridge/hag\-gle/imo\-te/cam\-bri\-d%
ge, jan 2006.

\bibitem{upmcCambridgeData}
J.~Leguay, A.~Lindgren, J.~Scott, T.~Riedman, J.~Crow\-croft, and P.~Hui,
  ``{CRAW\-DAD} trace upmc/content/imote/cambridge (v. 2006--11--17).''
  Downloaded from
  http://craw\-dad.cs.dart\-mouth.edu/\-upmc/con\-tent/imo\-te/cam\-bri\-dge,
  nov 2006.

\bibitem{cambridgeInfocomData}
J.~Scott, R.~Gass, J.~Crowcroft, P.~Hui, C.~Diot, and A.~Chaintreau,
  ``{CRAWDAD} trace cambridge/haggle/imote/infocom (v. 2006--01--31).''
  Downloaded from
  http://craw\-dad.cs.dart\-mouth.edu/cam\-bridge/hag\-gle/imote/infocom, jan
  2006.

\bibitem{cai08mobihoc}
H.~Cai and D.~Y. Eun, ``Toward stochastic anatomy of inter-meeting time
  distribution under general mobility models,'' in {\em MobiHoc '08:
  Proceedings of the 9th ACM international symposium on Mobile ad hoc
  networking and computing}, pp.~273--282, ACM, 2008.

\bibitem{hui07community}
P.~Hui, E.~Yoneki, S.~Y. Chan, J.~Crowcroft, and J.~Crowcroft, ``Distributed
  community detection in delay tolerant networks,'' in {\em MobiArch '07:
  Proceedings of first ACM/IEEE international workshop on Mobility in the
  evolving internet architecture}, pp.~1--8, ACM, 2007.

\bibitem{hui07socio}
E.~Yoneki, P.~Hui, S.~Chan, and J.~Crowcroft, ``A socio-aware overlay for
  publish/subscribe communication in delay tolerant networks,'' in {\em MSWiM
  '07: Proceedings of the 10th ACM Symposium on Modeling, analysis, and
  simulation of wireless and mobile systems}, pp.~225--234, ACM, 2007.

\bibitem{hui08mobihoc}
P.~Hui, J.~Crowcroft, and E.~Yoneki, ``Bubble rap: social-based forwarding in
  delay tolerant networks,'' in {\em MobiHoc '08: Proceedings of the 9th ACM
  international symposium on Mobile ad hoc networking and computing},
  pp.~241--250, ACM, 2008.

\end{thebibliography}
\end{document}